\documentclass[
]{ceurart}
\usepackage{xcolor}
\usepackage{todonotes}
\usepackage{sigirstyle}
\usepackage{setspace}

\renewcommand{\para}[1]{\textbf{#1}}

\usepackage{listings}
\begin{document}

\copyrightyear{2024}
\copyrightclause{Copyright for this paper by its authors.
  Use permitted under Creative Commons License Attribution 4.0
  International (CC BY 4.0).}

\conference{Woodstock'22: Symposium on the irreproducible science,
  June 07--11, 2022, Woodstock, NY}

\title{Top-Down Partitioning for Efficient List-Wise Ranking}

\author[1]{Andrew Parry}[%
orcid=0000-0001-5446-8328,
email=a.parry.1@research.gla.ac.uk,
url=https://parry-parry.github.io/,
]
\cormark[1]
\address[1]{University of Glasgow,
  Glasgow, UK}

\author[1]{Sean MacAvaney}[%
orcid=0000-0002-8914-2659,
email=Sean.MacAvaney@glasgow.ac.uk,
url=https://macavaney.us/,
]

\author[1]{Debasis Ganguly}[%
orcid=0000-0003-0050-7138,
email=Debasis.Ganguly@glasgow.ac.uk,
url=https://gdebasis.github.io/,
]

\cortext[1]{Corresponding author.}

\begin{abstract}
  Large Language Models (LLMs) have significantly impacted many facets of natural language processing and information retrieval. Unlike previous encoder-based approaches, the enlarged context window of these generative models allows for ranking multiple documents at once, commonly called list-wise ranking. However, there are still limits to the number of documents that can be ranked in a single inference of the model, leading to the broad adoption of a sliding window approach to identify the $k$ most relevant items in a ranked list. We argue that the sliding window approach is not well-suited for list-wise re-ranking because it (1) cannot be parallelized in its current form, (2) leads to redundant computational steps repeatedly re-scoring the best set of documents as it works its way up the initial ranking, and (3) prioritizes the lowest-ranked documents for scoring rather than the highest-ranked documents by taking a bottom-up approach. Motivated by these shortcomings and an initial study that shows list-wise rankers are biased towards relevant documents at the start of their context window, we propose a novel algorithm that partitions a ranking to depth $k$ and processes documents top-down. Unlike sliding window approaches, our algorithm is inherently parallelizable due to the use of a pivot element, which can be compared to documents down to an arbitrary depth concurrently. In doing so, we reduce the number of expected inference calls by around 33\% when ranking at depth 100 while matching the performance of prior approaches across multiple strong re-rankers.
\end{abstract}


\maketitle

\section{Introduction}

Until recently, neural approaches to ranking entailed the direct comparison of a document\footnote{In practice, these models are conducted over shorter passages of text. But for ease of reading, we simply refer to them as ``documents''.} to only one or two other distinct texts at a time: either a ``point-wise'' comparison to the query~\citep{dpr, monobert} or a ``pair-wise'' comparison to a query and one other reference document~\citep{expando}. Recent developments in Large Language Models (LLMs) improve the natural language understanding of transformer-based models and vastly increase the number of tokens a model can effectively process compared to previous models~\citep{flant5, gpt, llama}. These two improvements allow for a ``list-wise'' ranking approach at inference time, where multiple documents can all be evaluated simultaneously. The key benefit of such a list-wise approach is that the ranker is directly given context to judge each document by considering other potentially relevant documents. State-of-the-art list-wise ranking approaches output a permutation of the original ranking instead of explicit scores~\citep{rankgpt, rankzephyr}. The need to encode multiple documents and output multiple tokens to determine document order makes these models computationally expensive due to the quadratic complexity of attention, meaning that encoding multiple documents inherently takes more time than current point-wise architectures.

\begin{figure}[t]
    \centering
    \includegraphics[width=0.5\textwidth]{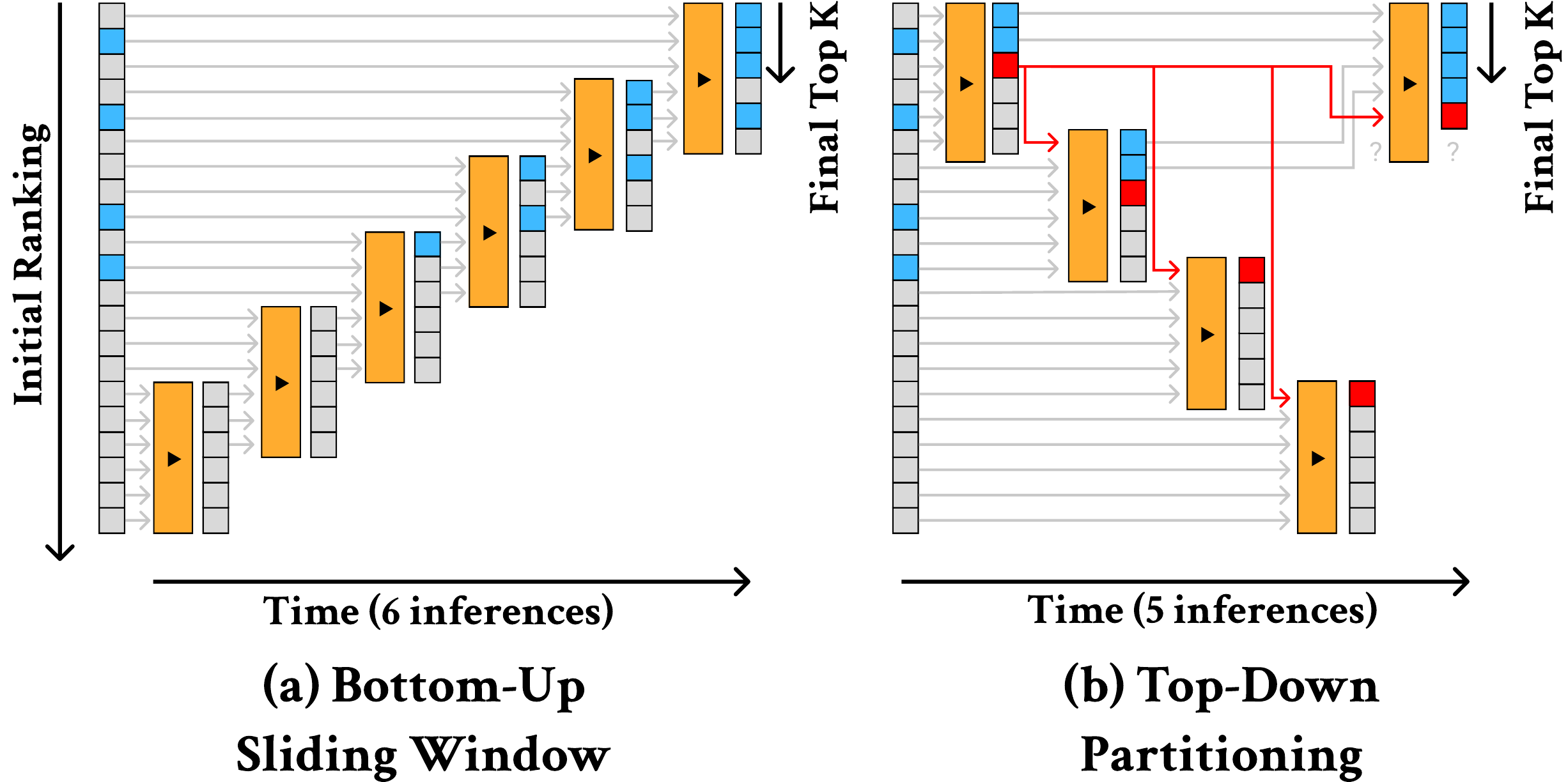}   
    \caption{Bottom-up sliding window vs our top-down partitioning approach. Relevant documents are colored blue. In the case of our approach, the pivot document marked red is identified in the first pass and can be compared to other partitions concurrently (Shown with red arrows).}
    \label{fig:compare}
\end{figure}

In existing work, there is still a limit to the number of documents that can be ranked in a single pass (typically around 20 documents~\citep{rankzephyr}). Therefore, a larger ranking must be `batched' before inference. A strategy is needed to select which passages to `batch' together and how to combine them, given that a batch permutation gives no indicator of absolute relevance across the full ranking---only the relative ranking of documents within a batch. Due to the absence of scores, this problem becomes closer to a classical selection or partitioning algorithm, such as quick select~\citep{quickselect}.

The prevailing approach in list-wise ranking uses a sliding window over a chosen ranking depth from bottom to top~\citep{rankgpt, rankwogpt, rankzephyr} as illustrated in Figure~\ref{fig:compare}(a). By using an overlapping `stride' over windows, documents that are initially lower in the ranking can bubble to the top, in principle~\citep{rankwogpt}. However, in this approach, the ranking of each window is dependent on the processing of the documents below, leading to the repeated scoring of multiple documents to enter the top-$k$, thus indicating that this is an inefficient partitioning scheme. We alleviate this problem with a rank-biased partitioning algorithm, which allows for parallel comparison of multiple batches since each batch only depends on a `pivot' from the first batch (Figure~\ref{fig:compare}(b)). Additionally, positional bias noted in prior work~\cite{permutation} may be amplified by the repeated re-scoring of a potentially relevant document to reach a high rank. Though prior work has investigated this effect under the effect of a sliding window~\cite{rankwogpt, schlatt2024setencoder}, we investigate the in-window precision of list-wise rankers over single inferences as opposed to ranked lists agnostic of a given partitioning algorithm to better understand their behaviour.

Our key intuition relates to the purpose of such expensive rankers; such an approach is inherently precision-orientated and, therefore, could find use in a broad set of domains that rely on either an exact match or a small set of highly relevant documents~\cite{rag, fid}. With such applications in mind, we identify a pivot element at the $k^{th}$ rank, which is then compared with all documents down to an arbitrary depth. Any document considered more relevant than the pivot element is necessarily a candidate for the top $k$ results. A final scoring is performed over all candidates to yield a final top-$k$. We additionally constrain the number of candidates that can be selected in a given iteration to allow a budget for latency. In improving both the computational expense and latency of list-wise rankers, we not only aim to improve their application at scale but make their use as annotators of training data more affordable as list-wise distillation becomes an increasingly common component of state-of-the-art ranking models~\cite{rankzephyr, rankt5, lassance2024spladev3}. Given the potential applications of these models, it is worthwhile to assess their efficiency and robustness; as such, we investigate the following research questions to verify the effectiveness of our approach.
\label{sec:rqs}
\textbf{RQ-1}: How does the proportion of relevance within a window change effectiveness when order changes?

We find that due to the list-wise nature of LLM-based rankers, they present a bias to list order even within a single context window. When the order of a ranking by human-judged relevance is inverted, a standard cross-encoder can consistently outperform a list-wise ranker. We generally observe a bias for a correctly ordered ranking, particularly in the case of a small proportion of relevant documents. Notably, we find that a standard cross-encoder achieves nearly identical performance to list-wise rankers over synthetic rankings when using common partition sizes.

\textbf{RQ-2}: What is the efficiency/effectiveness trade-off between different partitioning algorithms for list-wise ranking?  

Given our findings investigating \textbf{RQ-1}, we motivate our novel approach to list-wise ranking that matches the performance of sliding window ranking and reduces the number of LLM inferences while being parallelizable. We find that inferences can be reduced by up to 33\% with no significant loss in performance whilst additionally reducing the sequential dependence between windows found in prior approaches, allowing parallelism to improve efficiency from both a compute and latency perspective. Additionally, we find that though a single inference is sufficient when evaluating MSMARCO, other out-of-domain datasets can require ranking to a greater depth, where our algorithm can provide efficiency gains.

\textbf{RQ-3}: How sensitive is top-down partitioning to the precision of a first-stage retriever?

We find that our new approach can be sensitive when the initial window of documents is imprecise. This is due to our use of a pivot element, which can affect overall performance if poorly chosen. However, we can alleviate this effect due to the design of our algorithm, which allows for a budget over the number of candidate documents to collect, which can be increased in the case of a weak first-stage retriever.

\textbf{RQ-4}: How does a larger candidate pool affect ranking performance? 

In alleviating the effects of pivot element choice, we find that our algorithm successfully recovers from a sub-optimal pivot element when the allocated ranking budget is improved, increasing nDCG@10 by over two points. In doing so, we show that our algorithm can trade off efficiency for effectiveness given a weaker initial retrieved list.

With this work, we aim to bring attention to the inefficiencies of current approaches and how a task-specific approach can largely reduce the required inferences over computationally expensive language models. We release artifacts and experiment code to ensure reproducibility.\footnote{\href{https://github.com/Parry-Parry/TDPart}{Github Repository}}

\section{Background and Related Work}

\para{The Ranking Problem.}
A ranking is a permutation of a set ordered with respect to some scoring function. Within retrieval, for a corpus $C=\{d_1, \dots, d_{|C|}\}$ and a user query text $q$, a ranking model returns a top-$k$ set of documents where $k \ll |C|$, ordered by the probability of relevance to $q$~\citep{robertson1977probability}.

Until the resurgence of neural networks and, more specifically, the transformer architecture~\citep{transformer}, ad-hoc search primarily involved term weights from exact lexical matches~\citep{bm25}; neural architectures overcome the problem of exact term matching via soft-matching of terms using pre-trained language models~\citep{bert, t5}. Multiple variants of the transformer architecture are applied in text retrieval, with most approaches falling into one of the following two paradigms. The first is a bi-encoder, which separately encodes a query and document into a dense latent space and executes a vector similarity search to determine relevance~\citep{dpr, cedr}, a variation of this architecture, learned sparse retrieval uses sparse vectors frequently projected over the vocabulary~\citep{epic, splade}. The second, a cross-encoder, treats retrieval as a classification problem~\citep{monobert, monot5, monoelectra}, jointly encoding a query and document before outputting a relevance score for the pair. List-wise ranking can be seen as a cross-encoder variant due to its jointly encoding queries and multiple documents~\citep{rankgpt}. Though a list-wise loss criterion has been applied in Learning-To-Rank~\cite {listwiseltr} before neural retrieval, due to context window constraints, neural rankers were trained with point-wise or pair-wise loss functions. Recently, approaches using both T5~\cite{rankt5} and ELECTRA~\cite{schlatt2024setencoder} architectures have applied list-wise loss criteria as larger batch sizes can be accommodated in modern compute instances.

\para{List-wise Ranking.}
List-wise ranking is the ranking of multiple documents in a single inference step. We follow the notation convention of ~\citet{permutation} in which an $n$-permutation of an $n$-ranking $R=\{R_i\}_{i=1}^n$ is defined as $\sigma : \{d_1, \dots, d_n\} \rightarrow \{d_1, \dots, d_n\}$. Given $R$, a list-wise model $\theta$ outputs a permutation $\sigma$ ordered by relevance to $q$. A neural approach generally performs permutation in discrete space via the decoding of tokens to determine order. For an initial ranking $R$ and query text $q$, $\sigma(R) = \text{PERMUTE}(R, q;\theta)$, where the permutation $\sigma$ is generated auto-regressively through greedy decoding; however, our approach is not constrained only to methods applying autoregressive decoders. \citet{rankgpt} first showed the capability of GPT variants~\citep{gpt} as zero-shot list-wise rankers, with~\citet{lrl} achieving strong performance with a different prompt. Multiple works then distilled GPT-based rankers into smaller architectures in both a decoder-only~\citep{rankvicuna, rankzephyr} and encoder-decoder setting~\citep{lit5}, finding that the distilled model could outperform its teacher with suitable data augmentation~\citep{rankzephyr}. \citet{rankwogpt} then showed that comparable performance could be achieved without using the GPT API and additionally, investigated position bias across multiple windows; we instead investigate in-window precision to understand better the behaviour of list-wise rankers agnostic of a chosen partitioning algorithm.

Current approaches use sliding windows over a ranking to allow for inference in the case that the inclusion of all candidate documents would exceed the maximum context window of many models without excessive truncation. These methods chose a partition size $w$ where $w \ll |R|$ and progressively re-rank results from bottom to top by taking strides over windows. These methods effectively cannot be applied top to bottom as doing so provides an opportunity for highly ranked documents to be penalised as opposed to finding relevant documents that are ranked lower by a first-stage retriever, defeating the point of a cascading ranking pipeline. This naturally leads to inefficient search due to the need to sequentially search over each partition from bottom to top, particularly in the case of repeated re-ranking~\citep{rankzephyr}. Recent work has attempted to partially alleviate these inefficiencies in proposing the ``set-encoder'' architecture~\citep{schlatt2024setencoder}, which can process all documents in a single window; however, such an architecture requires annotated data from larger list-wise rankers which still employ a sliding window as it outperforms a full list-wise ranking, as such our work can not only improve efficiency in a search setting but allow for the efficient annotation of training data for more downstream ranking models. \citet{truncate} instead apply auxiliary models to dynamically adjust re-ranking depth conditioned on estimates of relevance to improve efficiency by re-ranking a reduced number of documents, where this approach is applied pre-inference, we instead employ a dynamic algorithm which allows early stopping in the case of a suitable number of relevant documents being found, however, our work is not independent of pre-inference approaches, and they could be combined to improve efficiency within this paradigm. Beyond list-wise approaches, \citet{pairwise} proposed a zero-shot pair-wise approach in which, much like a classic sorting algorithm, comparisons are made between elements to produce a ranked list. This parallel with sorting algorithms led to an expansion of this approach by \citet{setwiserank}, finding that efficiency and performance could be further improved by taking inspiration from common sorting algorithms. Due to the pair-wise nature of these approaches, under current architectures, such approaches are less efficient empirically; however, they remain compelling due to the absence of an expensive training stage.

\para{Dynamic Pruning.}
\label{sec:prune}
Our approach takes inspiration from classic dynamic pruning algorithms in retrieval. The key notion is that one does not need to consider documents that cannot enter the top-k during scoring. Such an idea has been applied to approximate a top-$k$ ranking over sparse indices in a document-at-a-time fashion~\citep{wand}. For a document $d$ and query $q$, the query-dependent score of $d$ can be expressed as:

\begin{equation}
    \mathcal{S}(d, q) = \sum_{t \in q \cap d} \alpha_t w(t, d)
    \label{eq:ub}
\end{equation}
where in Equation \ref{eq:ub}, $\alpha_t$ is the importance of a term $t$ within $q$ and $w(d, t)$ is the weight assigned to $t$ within $d$. For each query term the upper bound $\text{UB}_t$ of its contribution to $d$ is bounded by $\text{UB}_t \geq \alpha_t \ w(d, t)$, an upper bound on the query-dependent score of a document $d$ is then expressed as $\sum_{t \in q \cap d} \text{UB}_t \geq \mathcal{S}(d, q)$. If this bound is lower than the current $k^{\text{th}}$ document, the document cannot enter the top-$k$; as such, the scoring of many documents can be cut short or eliminated based on this estimation~\citep{mainmemretr}. In recent years, dynamic pruning has been successfully applied to neural approaches, including re-ranking~\citep{Leonhardt2023EfficientNR} and learned sparse retrieval~\citep{Qiao2023OptimizingGT}. What differentiates our algorithm for list-wise ranking is that we cannot make in-document or term-at-a-time comparisons and, therefore, use an entire document text as a threshold.

\section{Methodology}

\label{sec:method}

We now outline our methodology for investigating in-window precision for list-wise rankers and our proposed algorithm.

\subsection{Position and Relevance}
\label{sec:methodwindow}
When a window-based approach is employed, an imprecise ranking may lead to the scattered distribution of relevant documents contrary to the distillation data commonly used to train these list-wise rankers, as illustrated in Figure \ref{fig:compare}. Given the sensitivity of list-wise rankers to list order across a ranked list, we consider how effectiveness can be affected by both order and relevance distribution within a single context window. Due to this sensitivity, prior work has proposed the use of a cascading process in which a list is repeatedly re-ranked to compensate for the initial imprecision of a first-stage retriever~\cite{rankzephyr}. We look to investigate this behaviour agnostic of any given partitioning algorithm.

More concretely, we consider binary relevance for the purposes of this investigation, much like classic ad-hoc IR test collections. A ranked list $R$ of size $k$ can be composed with a ratio $r$ denoting the proportion of relevant documents within $R$. Given a set of relevant documents $D^+$ and non-relevant documents $D^-$, we choose $kr$ elements from $D^+$ and $(1-r)k$ elements from $D^-$. Under an order-insensitive ranker with oracle knowledge, it is clear that the measured performance of $R$ would be monotonically increasing with $r$ agnostic of the ordering of $R$. However, we posit that current list-wise rankers could show reduced robustness to ordering and that we can better understand the mechanism of current approaches over an imprecise initial ranking in investigating effectiveness under these conditions.  

\subsection{Proposed Algorithm}

Given the order sensitivity of existing approaches, we consider the need for a relevant document that initially lies at a low rank to be repeatedly re-ranked to reach the top of a ranking can have negative effects when re-ranking an imprecise list and furthermore clearly shows the inefficiency of a sliding window approach.

Our core idea is to remove the dependence of a sequence of partitions of a ranking on each other. Currently, under a sliding window $w$ with a stride $s$, dependencies prevent parallelism and other efficiency improvements due to the bottom-up nature of such an approach. To improve efficiency, we partition a ranking in a top-down fashion compared to a highly ranked pivot element $p$ to select candidates under a budget $b$ for each iteration. At a high level, consider the upper bounds described in Section \ref{sec:prune}; our approach instead uses a pivot \textit{document} $p$ at rank $k$ as a threshold as opposed to term-based thresholds. Agnostic of list-wise approaches, we assume that implicitly, a document $d$ is scored with respect to $q$. If $\mathcal{S}(q, d) < \mathcal{S}(q, p)$ then $d$ cannot enter the top-$k$ ranking. In practice, we directly use the permutation of a set of documents to determine the above inequality as opposed to exact scores, as if a document is ranked lower than $p$, it is considered less relevant by a model.

As outlined in Algorithm \ref{algo:pivot}, given a set of documents $A_{i-1} \subseteq R$ being the previous iteration or the full ranked list, we process the top-$w$ candidates $L$ and take a pivot element\footnote{we choose $k=\frac{w}{2}$ for comparison to the commonly used stride of 5.} $p$ at rank $k$ with all elements above the pivot added to a candidate set $A_i$ with maximum size $b$, which is modified over each iteration; all other elements are added to a backfill set $B$, which persists throughout execution to retain all documents which are estimated to be less relevant than the pivot. The remainder of the ranked list $R$ can be processed in parallel in partitions of size of $w-1$. 

\begin{algorithm}[t]
\small
\setstretch{0.9}
\DontPrintSemicolon
\KwIn{Candidate set: $A_{i-1}$, Backfill set: $B$, Query: $q$, Window size: $w$, Budget: $b$, Optimization cutoff: $c$, List-wise ranker: $\theta$}
\KwOut{$\sigma(A_{i-1})$: $A_{i-1}$ ordered by relevance to $q$}
\Begin{
    $L \gets \ \{d_1,\ldots,d_w, d \in A_{i-1}\}$ \tcp*[r]{First $w$ documents}
    
    $R \gets A_{i-1}-L$ \tcp*[r]{Remaining documents}
    
    $L \gets \text{PERMUTE}(L, q;\theta)$ \tcp*[r]{Order current partition}
    
    $p \gets L[k]$ \tcp*[r]{Take pivot element}

    $A_{i} \gets L[1:k]$
    
    $B \gets B \cup L[k+1:|L|]$

    \While {$|A_{i-1}| > 0$ and $|A_i| < b$}{
        $L \gets \{d_1,\ldots,d_{w-1}, d \in R\}$
        
        $R \gets R-L$ 
        
        $L \gets \text{PERMUTE}(p \cup L, q;\theta)$\;
        
        $A_{i} \gets A_{i} \cup L[1:p]$ \tcp*[r]{Elements $> p$}
        
        $B \gets B \cup L[p+1:|L|]$ \tcp*[r]{Elements $< p$}
    }

    \textbf{return} ($|A_{i}| = k-1 $)\hspace{0.5em}?\hspace{0.5em}$A_{i} \cup p \cup B\hspace{0.5em}:\hspace{0.5em}\text{pivot}(A_{i}) \cup p \cup B$ 
}
\caption{Top-Down Partitioning}
\label{algo:pivot}
\end{algorithm}

The pivot $p$ is prepended to the next set $L$, which is then ordered by model $\theta$ to find documents ranked above $p$. Any elements of $L$ ranked above $p$ are added to $A$, and the remainder is added to $B$. If no element in $R$ is greater than the current pivot, we know no element can enter the top-$k$; therefore, our top set is already sorted as $L$. If candidates are found, the algorithm is executed on $A_i$ constrained to budget size $b$ (Algorithm \ref{algo:pivot} Line 14). For our main evaluation, the candidate size equals the window size. Therefore, a single extra iteration is required after traversing a full ranked list. Our algorithm incorporates the desirable rank bias of favoring the top-ranked documents because it works top-down to select a candidate pool constrained by budget $b$. As a document at rank 20 is more likely to be relevant than a document at rank 100, when a set of $b$ candidates is found, we posit that there may be no need to search the remaining partitions.

For $|R| > w$, a sliding window approach always has a worst-case number of inferences, $\text{inferences }(R) = \frac{|R|}{s} - 1$ which in practise could not be improved via early stopping due to the bottom to top application of a sliding window. As top-down partitioning does not use a stride over a sliding window, our expectation over the number of inferences is a function of the desired search depth, the window size, and the budget $b$. Window size beyond the top-$w$ is reduced due to the need for the pivot element to be included in each inference. As we do not know how many documents will be considered more relevant than the pivot, identifying $p_i$ at the beginning of the $i^\text{th}$ iteration, we define the set of possible ranking iterations $A$ as follows. Given the $0^{\text{th}}$ element $A_0 = R$:
\begin{equation}
    %A_{n+1} = \{d | \ \text{rank}(d) < \text{rank}(p_{i}), \forall d \in A_{i}\}_{:b}
    A_i = \{A_{i} : |A_{i}| < b \wedge \forall d \in A_{i-1}, \ \text{rank}(d) < \text{rank}(p_{i-1})\}
    \label{eq:set}
\end{equation}
We can then estimate a worst-case number of inferences by taking the sum of inferences over all iterations defined in Equation \ref{eq:set}. Empirically, this does not occur as the algorithm either finds $b$ candidates before processing all partitions or the ranker is sufficiently precise to meet the condition $|A_{i}|=k-1$ early.
\begin{equation}
    \text{inferences }(R) = \sum_{i=0}^{|A|} 1 + \frac{|A_i|-w}{w-1}
    \label{eq:inference}
\end{equation}
Equation \ref{eq:inference} degenerates to $\text{inferences }(R) = 2 +\frac{|R|-w}{w-1}$ in the case of $b=w$. We need a single inference to find the initial pivot at each stage; we then have $|A_{i}|-w$ documents remaining to process with a reduced window size $w-1$ as we must account for the pivot element. We then require a final inference to order the set of $w$ candidates. Though both a sliding window and our algorithm are of $O(n)$ complexity, empirically, both early stopping and the lack of a stride lead to reduced computation.

\section{Evaluation}

We now describe the evaluation of top-down partitioning for list-wise re-ranking with respect to our research questions outlined in Section \ref{sec:rqs} and our observations from empirical evidence. 

\subsection{List-Wise Ranking Approaches.}
We compare approaches to ranking with sequence-to-sequence list-wise rankers. We do not consider variants of pairwise re-ranking~\citep{pairwise, setwiserank} as their efficiency and performance were found to be inferior to the following approaches~\citep{rankvicuna}.

\noindent \textbf{Single Window (Single)}: Re-ranks the top-$w$ documents from a first stage ranking where $w$ is window size. Assumes sufficient recall within the top-$w$ of a ranked list retrieved by a first-stage model. \\
\noindent \textbf{Sliding Window (Sliding)}: For a given window size $w$ and stride $s$ re-ranks documents to depth $D$ using a sliding window from bottom to top. \\
\noindent \textbf{Top-Down Partitioning (TDPart)}: Our approach outlined in Algorithm \ref{algo:pivot}. Performs an initial search over a top-$w$ window to find a pivot at cutoff $k$ before searching to depth $D$ comparing each window to the pivot. 

\subsection{Datasets}
We conduct our core experiments on the MSMARCO corpus~\citep{msmarco}, a collection of over 8.8 million documents commonly used to train neural ranking models. As we are primarily concerned with the precision of the top-$k$ re-ranked documents, we choose to evaluate our approach on the TREC Deep Learning 2019 and 2020 test queries~\citep{dl19, dl20}, which have densely annotated queries in contrast to the official MSMARCO DEV set~\citep{msmarco}, we refer the reader to the original model papers for an overview of each models performance on sparse benchmarks. To evaluate the robustness of our approach in an out-of-domain setting we take a subset of the BEIR benchmark sets~\citep{beir}, due to the greater computational expense of current list-wise rankers multiple sparsely annotated test sets would be infeasible under our compute budget, as such we choose test sets with a smaller set of densely annotated topics namely, TREC Covid~\citep{trecovid} and Touche~\citep{touche}. TREC COVID is a test benchmark with a corpus of 171k texts and 50 annotated queries with topics focusing on health information relating to the COVID-19 virus. Touche is an argument retrieval benchmark with a corpus of 382k texts and 49 annotated queries with topics relating to both societally important and everyday information needs that require arguments to provide evidence for a given perspective. 

We evaluate nDCG@1, 5, 10, and P@10 using the ir\_measures library; following standard MSMARCO conventions, we consider a relevance judgment of 2 or more as relevant and a relevance judgement of 1 for other datasets. Our algorithm aims to reduce the inferences required when applying list-wise rankers while maintaining effectiveness, so we conduct equivalence tests (paired TOST) with $p<0.05$ and a 5\% bounds between our methods and baselines. Additionally, we report the average number of inferences required per query for in- and out-of-domain experiments and the number of possible parallel inferences.

\subsection{Models}
We apply our approach to multiple state-of-the-art list-wise rankers that execute the PERMUTE function described in Section \ref{sec:method}. RankGPT~\citep{rankgpt}, which directly prompts GPT variants\footnote{We use GPT 3.5 (\texttt{gpt-3.5-turbo-0125}) due to budget constraints}, a decoder-only distilled ranker RankZephyr~\citep{rankzephyr} based on the Zephyr-7B~\citep{tunstall2023zephyr} architecture and LiT5~\citep{lit5} based on a T5~\citep{t5} fusion-in-decoding model~\citep{fid}. We allow a maximum of 4096 tokens per input in all cases. We also employ an oracle approach with access to test relevance judgments returning a permutation sorted by these judgments to ensure our algorithm can remain robust as underlying models improve. Due to tie breaks inherent when sorting by discrete relevance judgments, precision can vary.  As our approach is fundamentally rank-biased, we investigate the effect of first-stage rankers on downstream effectiveness. We re-rank BM25~\citep{bm25}, a common lexical model, and we use the Terrier implementation with default parameters~\citep{terrier}; SPLADE++ ED~\citep{splade}, a teacher-distilled learned sparse model and RetroMAE~\citep{retromae}, a teacher-distilled bi-encoder using MAE pre-training over a brute force index. In experiments addressing \textbf{RQ-1}, we additionally contrast RankZephyr with monoELECTRA~\cite{monoelectra}, a strong point-wise cross-encoder to control for order-sensitivity in our comparisons.

\subsection{In-Window Effectiveness}

As described in Section \ref{sec:methodwindow}, we investigate how ranking order and relevance distribution affect the in-window effectiveness of list-wise rankers. In investigating \textbf{RQ-1}, we define a ranking data generation process with oracle knowledge, which generates ranked lists given a ratio of relevant and non-relevant documents. To reduce the noise in this process, given the existence of false negatives in MSMARCO, we persist in a ranking between ratios within the generation process, only adding relevant documents as the ratio $r$ increases. Using human relevance judgements for a test collection, we partition all judged documents $D$ into relevant, $D^+$, and non-relevant, $D^-$, sets. We filter queries\footnote{We use the same filtered set of queries for different window sizes to allow comparisons, i.e., both window sizes sample from the pool satisfying $|R|=20$.} such that all evaluated queries have at least $k-1$ judgements in both $D^+$ and $D^-$ for $|R|=k$ to allow for the smooth interpolation between majority relevant and majority non-relevant ranked lists. For experiments with MSMARCO, we combine Deep Learning 2019 and 2020 queries, yielding 53 compatible queries for $|R|=20$. For out-of-domain experiments, 50 TREC COVID queries and 16 Touche queries are valid. Naturally, we use the full set of queries when performing retrieval effectiveness evaluation outside of this synthetic setting.

To persist a ranked list, for an initial ratio $r_0$, we take documents from the sets described above. When generating for the next ratio $r_1, \text{where} \ r_1 > r_0$, we find the number of \textit{new} relevant documents we need to include via $n=(r_1-r_0)|R|$. We remove $n$ non-relevant documents from $R$ and add $n$ documents from $D^+$, ensuring that random noise is reduced and we can clearly measure the effect of adding new relevant documents across a given sample.

To investigate order-sensitivity, we again utilise oracle knowledge to sort by relevance judgement; on MSMARCO, this is a value in the range $[0, 1, 2, 3]$ and generally, for BEIR datasets, a value in the range $[0, 1, 2]$. We sort each ranking in ascending (\textbf{ASC}), descending (\textbf{DESC}), and random order (\textbf{RANDOM}). In doing so, we investigate how the behaviour of these models varies with lists, which could be present at various points in the re-ranking process. For example, when re-ranking a strong first-stage, the top-k is likely to be well ordered, approximating descending order with a majority of relevant documents present; however, when re-ranking a weaker first stage, when ranking lower ranked documents, a window may be composed of only a few relevant documents in either random or ascending order. As such, our synthetic conditions can have parallels with real ranked lists.

To prevent bias in the initialisation of each ranked list, we aggregate performance over 5 initial rankings for each query. We investigate rations in the range $[0.2, 0.4, 0.6, 0.8]$, where a smaller ratio denotes a reduced proportion of relevant documents. We assess window sizes of 5 and 20 to observe how these conditions vary when a reduced number of documents are observed. 

\section{Results and Discussion}

We now present and discuss our findings with respect to research questions outlined in Section \ref{sec:rqs}.

\subsection{In-Window Effectiveness}

\begin{figure}
    \centering
    \subfloat[MSMARCO ($w=5$)]{\includegraphics[width=0.33\textwidth]{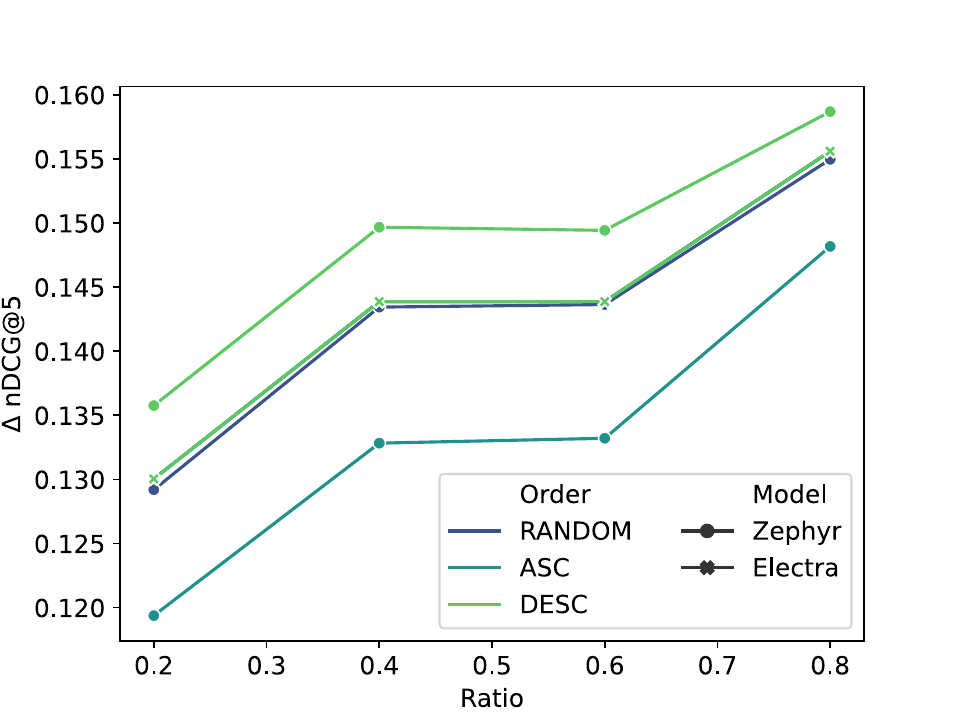}}\hfill
    \subfloat[ToucheV2 ($w=5$)]{\includegraphics[width=0.33\textwidth]{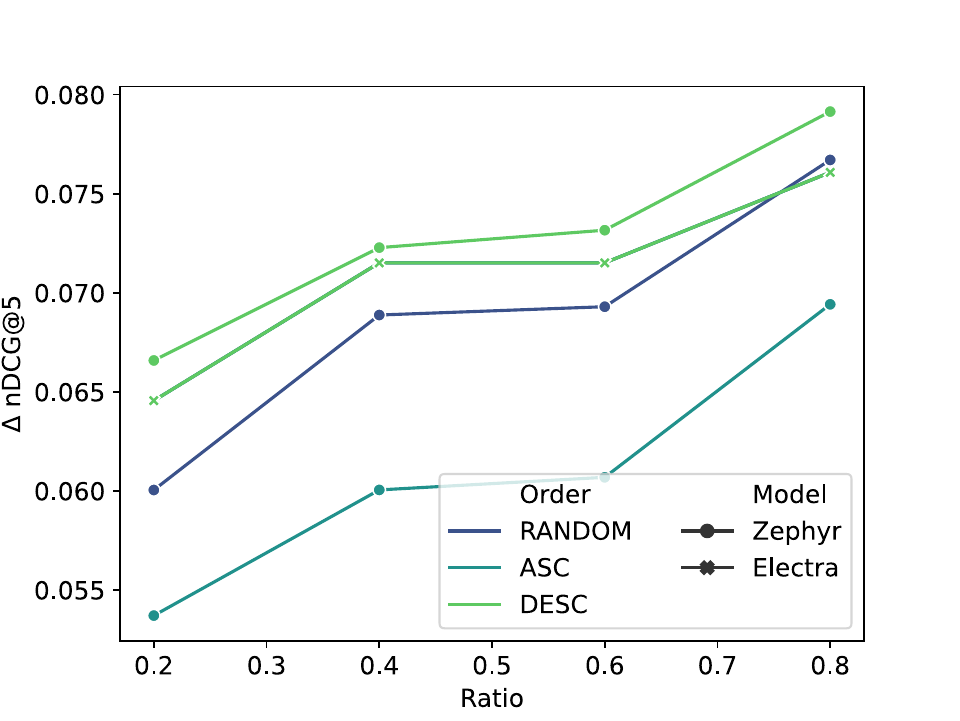}}\hfill
     \subfloat[COVID ($w=5$)]{\includegraphics[width=0.33\textwidth]{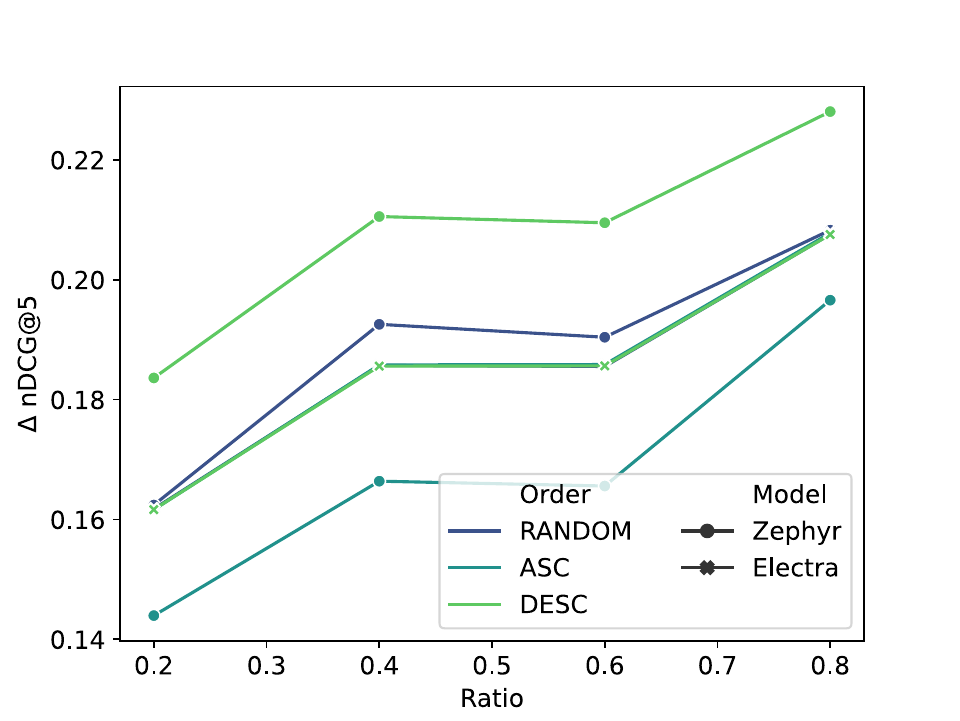}}\hfill
    \subfloat[MSMARCO ($w=20$)]{\includegraphics[width=0.33\textwidth]{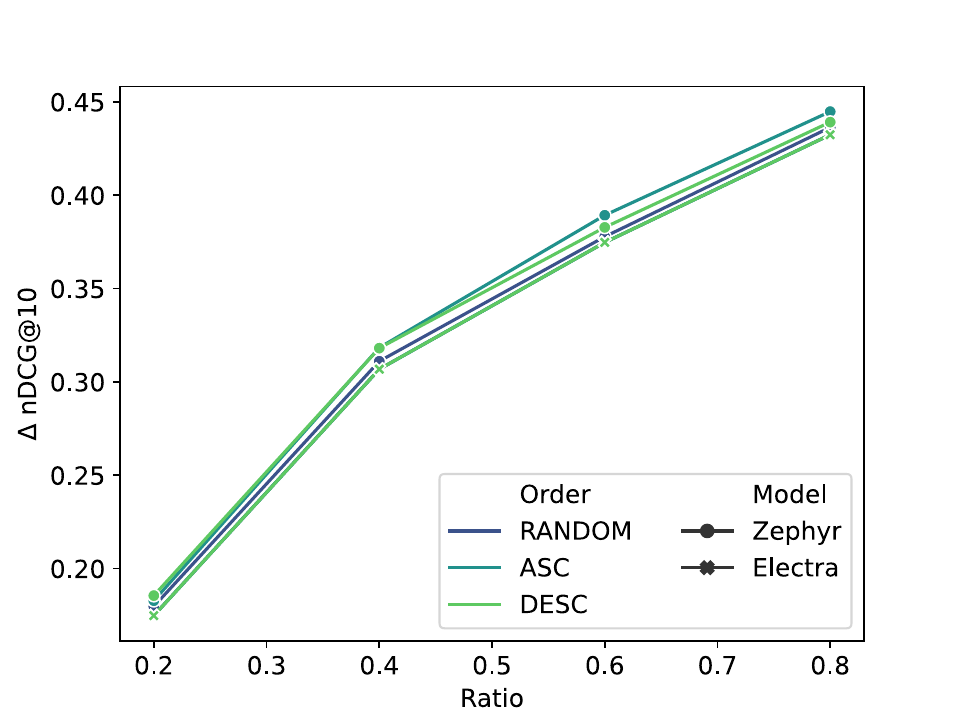}}\hfill
    \subfloat[ToucheV2 ($w=20$)]{\includegraphics[width=0.33\textwidth]{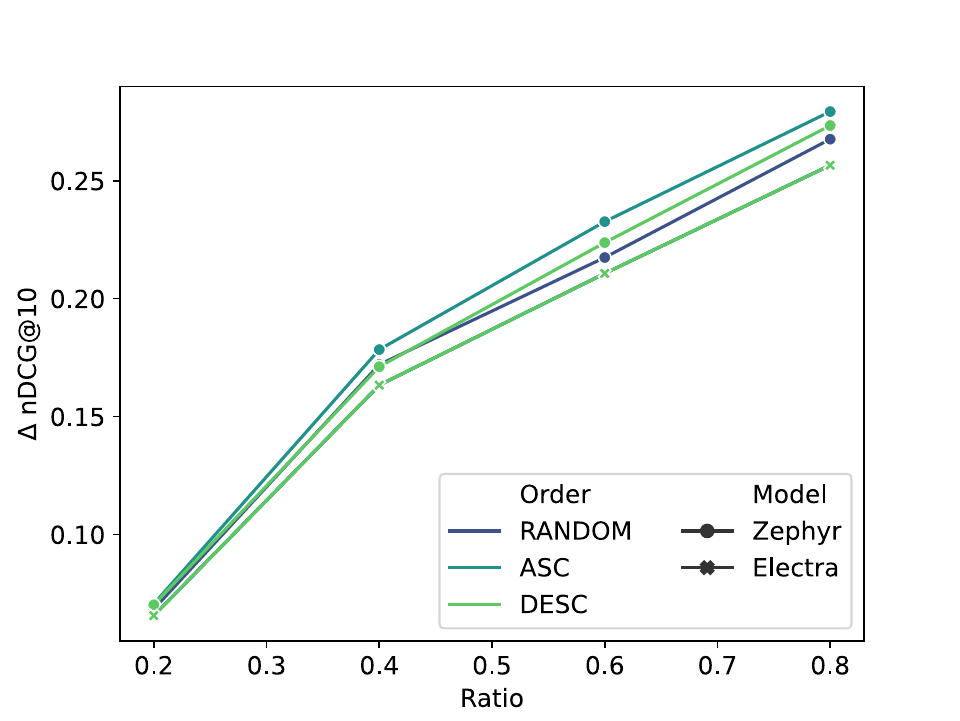}}\hfill
    \subfloat[COVID ($w=20$)]{\includegraphics[width=0.33\textwidth]{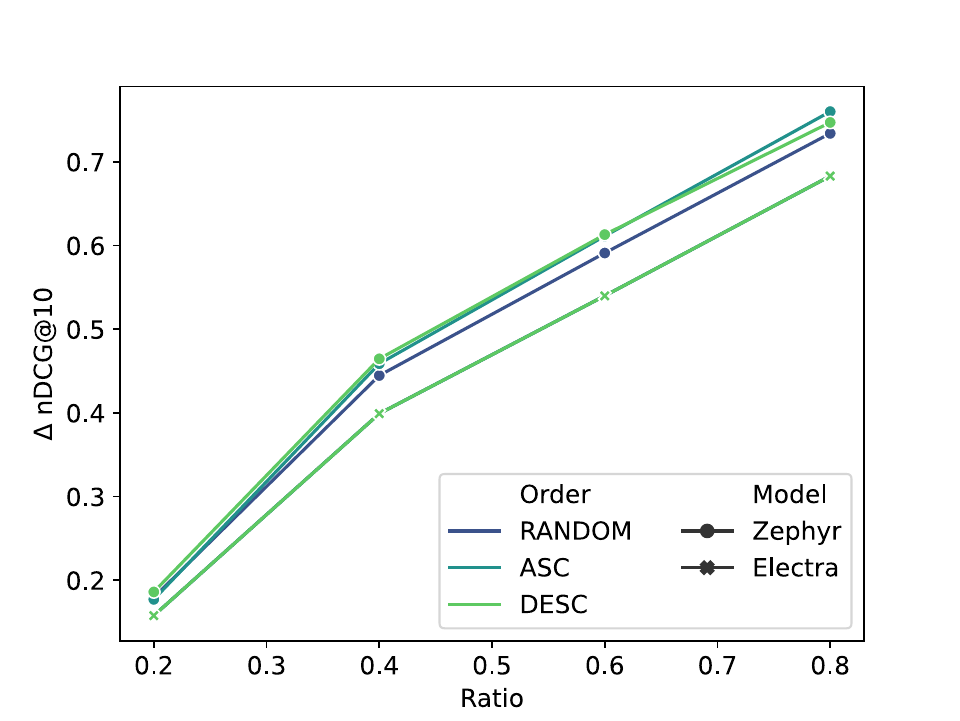}}
    \caption{Effectiveness of a list-wise and point-wise ranker varying the proportion of relevant documents and the ranked order measuring nDCG@10 across MSMARCO Deep Learning, Touche and TREC COVID queries. $w$ represents window size or, in this case, the number of documents ranked at once.}
    \label{fig:ratio}
\end{figure}
\begin{comment}
\begin{figure}[t]
    \centering
    \includegraphics[width=\textwidth]{src/figure/runthrough.pdf}   
    \caption{Illustration of RankZephyr top-down partitioning over a SPLADE retrieved Deep Learning 2019 query. Stage 1 selects a pivot element at a cutoff and subsequent stages make comparisons to this pivot. After no more candidates are found or a budget is exceeded, the final stage re-ranks all candidates.}
    \label{fig:execute}
\end{figure}
\end{comment}
In Figure \ref{fig:ratio}, we present the effectiveness of RankZephyr\footnote{Similar trends were consistently observed for other list-wise rankers} compared to monoELECTRA (order-invariant), controlling for the proportion of relevant documents (Ratio) and the ordering of the ranking. Addressing \textbf{RQ-1}, Figures \ref{fig:ratio} (a), (b), and (c) indicate that for a reduced list length, there is a clear distinction between orderings suggesting a greater sensitivity to relevant documents being placed at the beginning of a ranking. Additionally, observe that RankZephyr, when re-ranking from a ranked list in descending order, consistently outperforms monoELECTRA; however, random order is aligned with monoELECTRA in-domain, and performance varies out-of-domain. 

In contrast, we observe that in Figures \ref{fig:ratio}, RankZephyr exhibits minimal order sensitivity. Additionally, it is notable that monoELECTRA performs near-identically to RankZephyr under this synthetic setting. Given that a greater margin is observed between list-wise and point-wise methods out-of-domain, as reported by \citet{schlatt2024setencoder}, we consider that ``inter-document" interactions may be unnecessary when parameters are well-modelled in-domain. List-wise inference provides appropriate context when the target domain is not fully captured in a point-wise setting; however, clearly, list-wise distillation can improve performance, but list-wise inference may lead to a bias to rankings similar to that of a teacher model, as discussed below in Section \ref{sec:marco}.

Coupled with this finding, we observe that descending order is preferred in both in- and out-of-domain settings for lower proportions of relevant documents, suggesting that a well-ordered ranking can further improve performance. We draw a parallel with repeated re-ranking over BM25 proposed by \citet{rankzephyr}, in which initially, a relevant document may be missed as it may happen to be in a low position in the ranking and its rank is not sufficiently increased in the current sliding window. However, when re-ranking is again performed, the position of that document may have been moved such that the document rank is sufficiently improved to consider it at an earlier position in the context window, aligning with descending order for a low ratio as discussed above\footnote{This observation may be partially attributed to causal attention, which, in the case of descending order, the model can contextualise non-relevant documents with respect to already observed relevant documents however we leave verification of this conjecture to future work.}. 

When applying top-down partitioning, we consider that this effect may be reduced by placing the pivot element used for candidate selection at the first position in the ranking. Assuming that a ranked list is suitably well-ordered to allow the chosen pivot to be partially relevant, this should align with our findings that descending order is preferred for a reduced number of relevant documents.

\subsection{In-Domain Effectiveness}
\begin{table}[htp]
\caption{Comparing `Single Window', `Sliding Window', and `TDPart' (ours) approaches on TREC Deep Learning benchmarks. We report TOST (p<0.05), equivalence to our approach denoted with \underline{underline}. We report mean inferences with the mean number of inferences run in parallel in parentheses.
}
\begin{adjustbox}{width=\textwidth}
\begin{tabular}{@{}lllrrrrrrrrr@{}}
\toprule
& & & \multicolumn{2}{c}{nDCG@1} & \multicolumn{2}{c}{nDCG@5} & \multicolumn{2}{c}{nDCG@10} & \multicolumn{2}{c}{P@10} &  \\
\cmidrule{4-12}
& Model & Mode & DL19 & DL20 & DL19 & DL20 & DL19 & DL20 & DL19 & DL20 & N. Inf.  \\
\midrule
\multirow{13}{*}{\rotatebox[origin=c]{90}{\textbf{SPLADE ++ ED}}} & Oracle & Single & \underline{0.977} & \underline{1.000} & 0.936 & \underline{0.977} & 0.890 & 0.916 & 0.765 & 0.748 & 1.0 (1.0) \\
& Oracle & Sliding & \underline{0.984} & \underline{1.000} & \underline{0.972} & \underline{0.995} & \underline{0.957} & \underline{0.978} & \underline{0.872} & \underline{0.822} & 9.0 (1.0) \\
& Oracle & TDPart & 0.984 & 1.000 & 0.972 & 0.995 & 0.956 & 0.976 & 0.872 & 0.820 & 7.0 (5.0) \\
\cmidrule{2-12}
& Zephyr & Single & \underline{0.849} & \underline{0.855} & \underline{0.808} & \underline{0.831} & \underline{0.777} & \underline{0.795} & \underline{0.672} & \underline{0.637} & 1.0 (1.0)  \\
& Zephyr & Sliding & \underline{0.849} & \underline{0.849} & \underline{0.791} & \underline{0.826} & \underline{0.777} & \underline{0.802} & \underline{0.691} & \underline{0.631} & 9.0 (1.0) \\
& Zephyr & TDPart & 0.833 & 0.852 & 0.794 & 0.828 & 0.780 & 0.805 & 0.679 & 0.639 & 7.4 (5.4)  \\
\cmidrule{2-12}
& LiT5 & Single & \underline{0.802} & \underline{0.827} & \underline{0.795} & \underline{0.797} & \underline{0.763} & \underline{0.763} & \underline{0.665} & \underline{0.598} & 1.0 (1.0) \\
& LiT5 & Sliding & \underline{0.802} & \underline{0.818} & \underline{0.798} & \underline{0.771} & \underline{0.774} & \underline{0.763} & \underline{0.691} & \underline{0.600} & 9.0 (1.0) \\
& LiT5 & TDPart & 0.787 & 0.827 & 0.795 & 0.791 & 0.766 & 0.768 & 0.684 & 0.604 & 6.3 (4.3) \\
\cmidrule{2-12}
& GPT 3.5 & Single & 0.833 & 0.787 & \underline{0.783} & \underline{0.775} & \underline{0.760} & \underline{0.752} & \underline{0.644} & 0.607 & 1.0 (1.0) \\
& GPT 3.5 & Sliding & 0.837 & 0.781 & \underline{0.786} & 0.779 & \underline{0.754} & 0.766 & \underline{0.642} & 0.617 & 9.0 (1.0)  \\
& GPT 3.5 & TDPart & 0.802 & 0.806 & 0.779 & 0.788 & 0.753 & 0.752 & 0.642 & 0.598 & 7.4 (5.4)  \\
\midrule
\multirow{13}{*}{\rotatebox[origin=c]{90}{\textbf{RetroMAE}}} & Oracle & Single & 0.953 & 0.981 & 0.925 & 0.934 & 0.863 & 0.874 & 0.749 & 0.696 & 1.0 (1.0) \\
& Oracle & Sliding & \underline{0.992} & \underline{1.000} & \underline{0.968} & \underline{0.993} & \underline{0.948} & \underline{0.958} & \underline{0.863} & \underline{0.800} & 9.0 (1.0) \\
& Oracle & TDPart & 0.992 & 1.000 & 0.967 & 0.993 & 0.945 & 0.958 & 0.853 & 0.800 & 7.1 (5.1) \\
\cmidrule{2-12}
& Zephyr & Single & \underline{0.810} & \underline{0.867} & \underline{0.792} & \underline{0.815} & \underline{0.758} & \underline{0.778} & 0.665 & \underline{0.607} & 1.0 (1.0) \\
& Zephyr & Sliding & \underline{0.810} & \underline{0.867} & \underline{0.796} & \underline{0.818} & \underline{0.769} & \underline{0.785} & \underline{0.686} & \underline{0.607} & 9.0 (1.0) \\
& Zephyr & TDPart & 0.826 & 0.864 & 0.797 & 0.814 & 0.771 & 0.788 & 0.686 & 0.619 & 7.2 (5.2) \\
\cmidrule{2-12}
& LiT5 & Single & 0.787 & \underline{0.858} & \underline{0.773} & \underline{0.776} & \underline{0.739} & \underline{0.748} & 0.653 & \underline{0.585} & 1.0 (1.0) \\
& LiT5 & Sliding & 0.756 & \underline{0.855} & \underline{0.770} & \underline{0.781} & \underline{0.741} & \underline{0.750} & 0.665 & \underline{0.578} & 9.0 (1.0)  \\
& LiT5 & TDPart & 0.740 & 0.867 & 0.763 & 0.798 & 0.743 & 0.761 & 0.684 & 0.591 & 6.6 (4.6) \\
\cmidrule{2-12}
& GPT 3.5 & Single & 0.841 & 0.827 & \underline{0.802} & 0.789 & \underline{0.741} & \underline{0.744} & 0.630 & \underline{0.576} & 1.0 (1.0) \\
& GPT 3.5 & Sliding & 0.802 & \underline{0.799} & 0.779 & \underline{0.752} & \underline{0.740} & \underline{0.727} & 0.649 & 0.561 & 9.0 (1.0) \\
& GPT 3.5 & TDPart & 0.837 & 0.796 & 0.787 & 0.758 & 0.736 & 0.739 & 0.637 & 0.578 & 7.2 (5.2) \\
\midrule
\multirow{13}{*}{\rotatebox[origin=c]{90}{\textbf{BM25}}} & Oracle & Single & 0.907 & 0.935 & 0.823 & 0.811 & 0.719 & 0.715 & 0.560 & 0.489 & 1.0 (1.0) \\
& Oracle & Sliding & \underline{0.953} & \underline{0.981} & \underline{0.924} & \underline{0.928} & \underline{0.879} & \underline{0.880} & \underline{0.788} & \underline{0.702} & 8.9 (1.0) \\
& Oracle & TDPart & 0.946 & 0.981 & 0.913 & 0.925 & 0.858 & 0.872 & 0.765 & 0.687 & 7.4 (5.4) \\
\cmidrule{2-12}
& Zephyr & Single & 0.713 & 0.784 & 0.681 & 0.689 & 0.625 & 0.628 & 0.500 & 0.435 & 1.0 (1.0) \\
& Zephyr & Sliding & 0.713 & \underline{0.846} & 0.715 & \underline{0.770} & 0.707 & \underline{0.722} & 0.637 & \underline{0.535} & 8.9 (1.0) \\
& Zephyr & TDPart & 0.736 & 0.858 & 0.731 & 0.772 & 0.681 & 0.723 & 0.577 & 0.543 & 7.4 (5.4) \\
\cmidrule{2-12}
& LiT5 & Single & 0.748 & 0.799 & 0.685 & 0.678 & 0.626 & 0.604 & 0.507 & 0.413 & 1.0 (1.0) \\
& LiT5 & Sliding & 0.787 & \underline{0.833} & 0.751 & \underline{0.735} & 0.723 & \underline{0.700} & 0.640 & 0.531 & 8.9 (1.0)  \\
& LiT5 & TDPart & 0.787 & 0.836 & 0.724 & 0.746 & 0.687 & 0.679 & 0.593 & 0.489 & 5.9 (3.9) \\
\cmidrule{2-12}
& GPT 3.5 & Single & 0.698 & 0.765 & 0.647 & 0.666 & 0.602 & 0.588 & 0.484 & 0.393 & 1.0 (1.0) \\
& GPT 3.5 & Sliding & 0.709 & 0.809 & 0.727 & \underline{0.704} & 0.686 & 0.654 & 0.588 & 0.480 & 8.9 (1.0) \\
& GPT 3.5 & TDPart & 0.686 & 0.802 & 0.693 & 0.708 & 0.654 & 0.637 & 0.556 & 0.459 & 7.5 (5.5) \\
\bottomrule
\end{tabular}
\end{adjustbox}
\label{tab:main}
\end{table}
\para{Inferences are Empirically Reduced.}
Table \ref{tab:main} presents our core results over TREC Deep Learning queries, contrasting our algorithm with existing approaches. Rows 1-3 show an oracle ranker over each first stage. We observe statistical equivalence with a sliding window approach while consistently reducing the number of inferences required to achieve this performance. In doing so, we show that the comparison of a pivot element as opposed to a strided window is sufficient for list-wise inference.

Addressing \textbf{RQ-2}, as can be observed in the last column of Table \ref{tab:main}, when applying state-of-the-art rankers, we again observe a reduced number of inferences across all cases when applying top-down partitioning, as expected by the analysis of our algorithm. Additionally, our approach exceeds or matches a sliding window's performance measured by nDCG@10 in 79\% of cases, including RankGPT, which is zero-shot and known to have variable output~\cite{rankvicuna}. Additionally, observe that under an oracle setting, our algorithm is more efficient depending on the precision of a first-stage retriever, showing that we can improve efficiency agnostic of the effectiveness of an underlying estimator. When utilising both decoder-only and encoder-decoder architectures, the number of inferences required does not show a linear trend, which could be attributed to the ability of each model to suitably identify relevant documents given their position within a current window; however, in each case, we see that our algorithm leads to at maximum .4 more inferences than an oracle setting on average per query. Crucially, the latency of our approach would be dramatically reduced due to the majority of inferences being parallelizable, as observed in the bracket of the final column of each Table. Particularly in the case of LiT5, which is based on the smaller T5 architecture, this could lead to greater potential for list-wise inference to scale under a greater number of concurrent queries. Due to the reduced window size required by our algorithm, we frequently observe a smaller final window as in our experiments, depth 100 does not cleanly divide by 19 ($w-1$); simply searching to an approximate depth which has the factor $w-1$ would further reduce inferences with minimal effect on performance.

\para{Sensitivity to Initial Ranking.}
\label{sec:marco}
Addressing \textbf{RQ-3}, when applying our algorithm to strong first stages such as RetroMAE and SPLADE++ ED, it can be observed in Table~\ref{tab:main} that in all cases, our approach is either statistically equivalent or outperforms the strongest baseline, being either a single window in the case that the top-$w$ documents are already sufficiently precise or a sliding window. Additionally, our approach generally outperforms a sliding window in terms of nDCG@5; however, we note that nDCG@1 is variable. We generally achieve statistical equivalence with a sliding window. This observation shows that our approach provides an effective trade-off between precision and latency whilst outperforming a single window where a sliding window fails. Of note is the statistical equivalence of all approaches when using a strong first stage and list-wise ranker; however, when considering a list-wise ranker, one would want to maximize performance such that a deeper search is essential. We propose that the multiple scoring of documents in a sliding window approach, which can `miss' relevant documents due to the need for multiple scoring of the same document as observed by \citet{rankwogpt} which may inversely pull non-relevant documents into the top-$k$, explaining performance degradation over a single window as can be observed particularly under RankZephyr. Alternatively, when applying a weaker first stage, such as BM25, though our approach improves over a single window, we observe non-equivalent performance in P@10 when applying our approach versus a sliding window. A likely cause of this degradation is that due to the imprecision of the first $w$ elements of a BM25 ranking, the chosen pivot element is insufficiently relevant to distinguish suitable candidates. Therefore, the candidate pool is polluted with partially or non-relevant items as the pivot is not sufficiently relevant to provide a challenging threshold. However, our approach more robustly ranks documents in contrast to a sliding window, as shown by improvements in nDCG@1 and nDCG@5, providing further evidence that non-relevant documents are accidentally `bubbled' up to high ranks. Applying a list-wise ranker with our approach saves computation; therefore, allocating a small fraction of these savings to a stronger first stage, such as SPLADE++ ED, is worthwhile. 

\begin{figure}
    \centering
    \subfloat[SPLADE++ ED]{\includegraphics[width=0.33\textwidth]{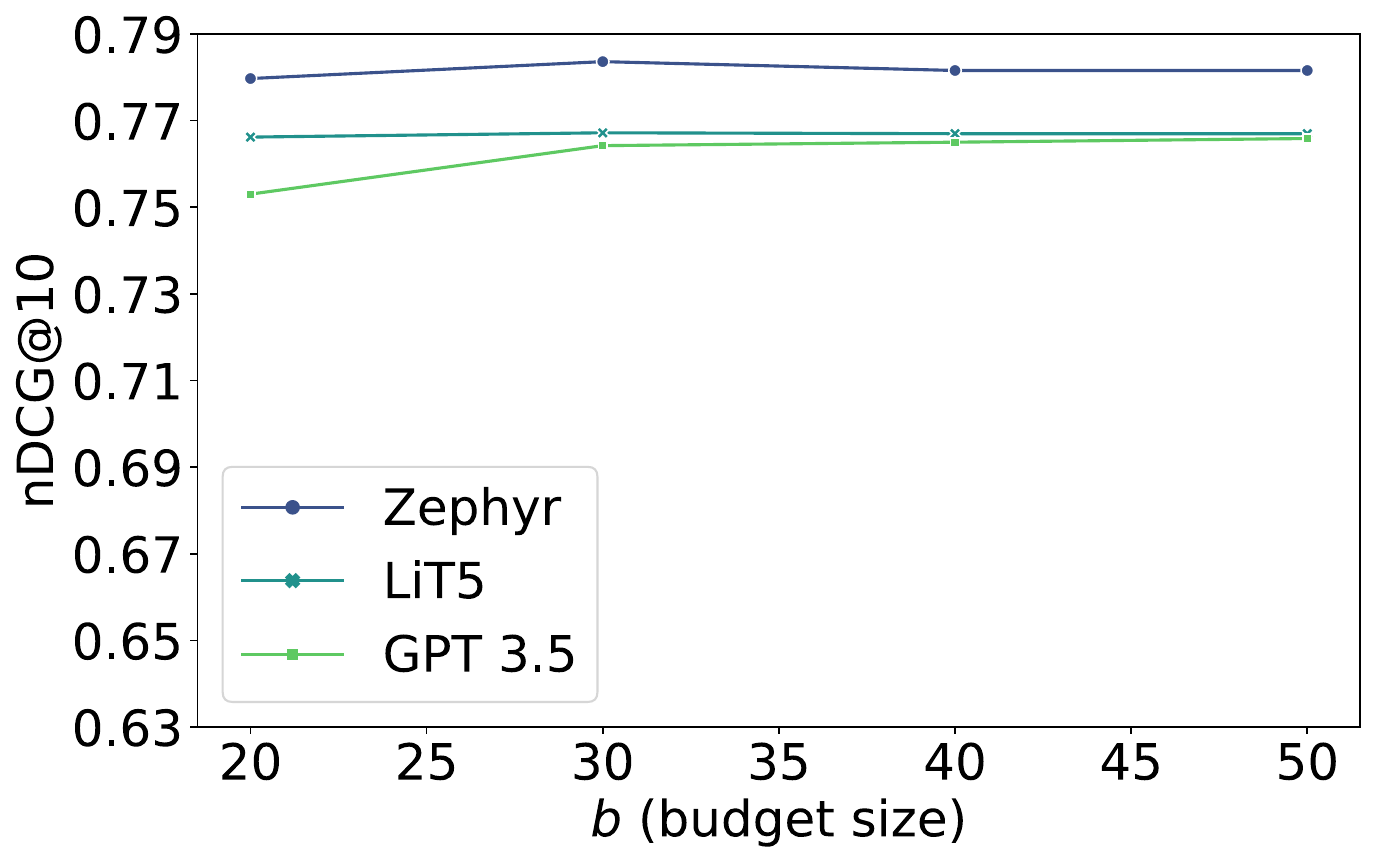}}\hfill
    \subfloat[RetroMAE]{\includegraphics[width=0.33\textwidth]{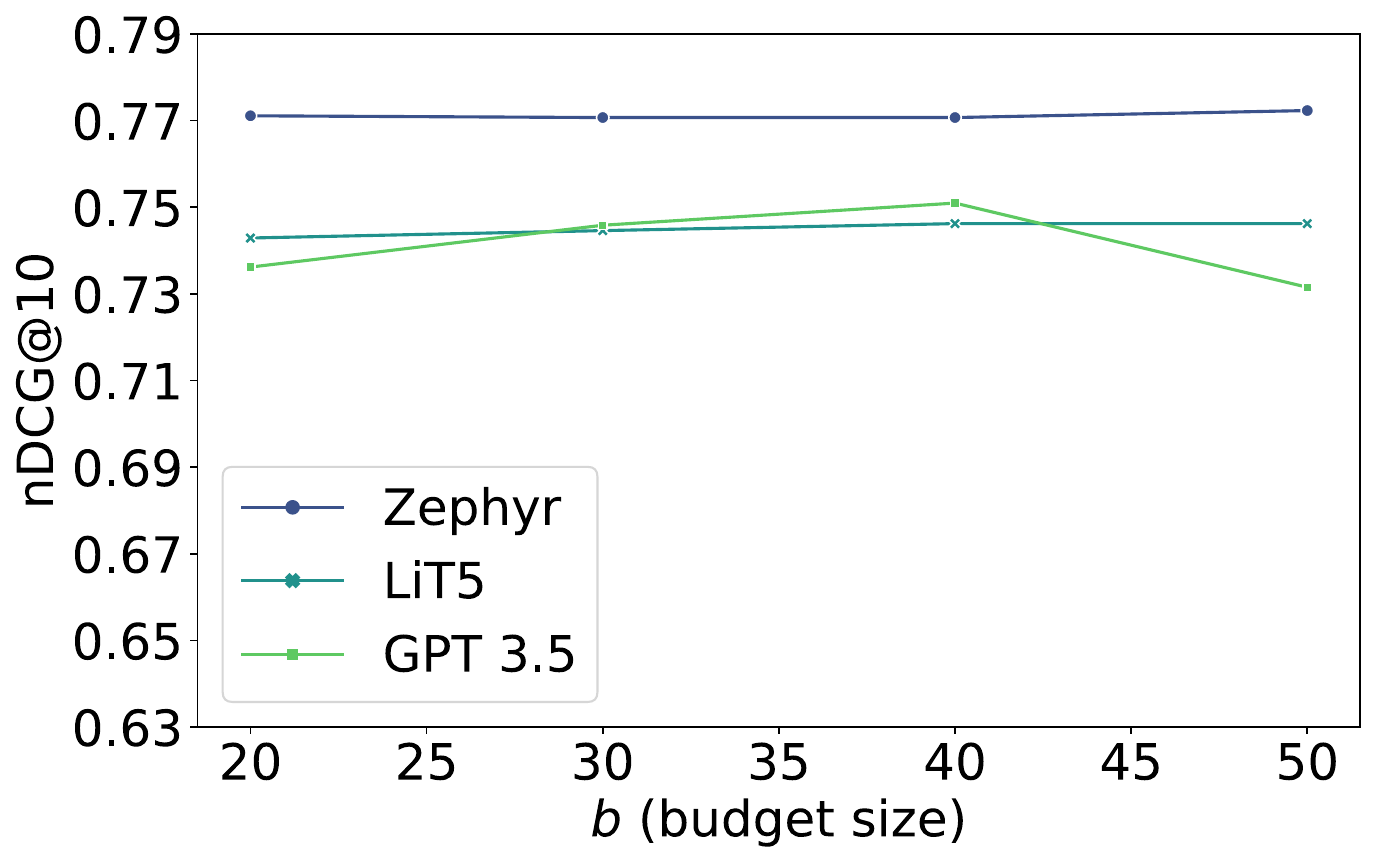}}\hfill
    \subfloat[BM25]{\includegraphics[width=0.33\textwidth]{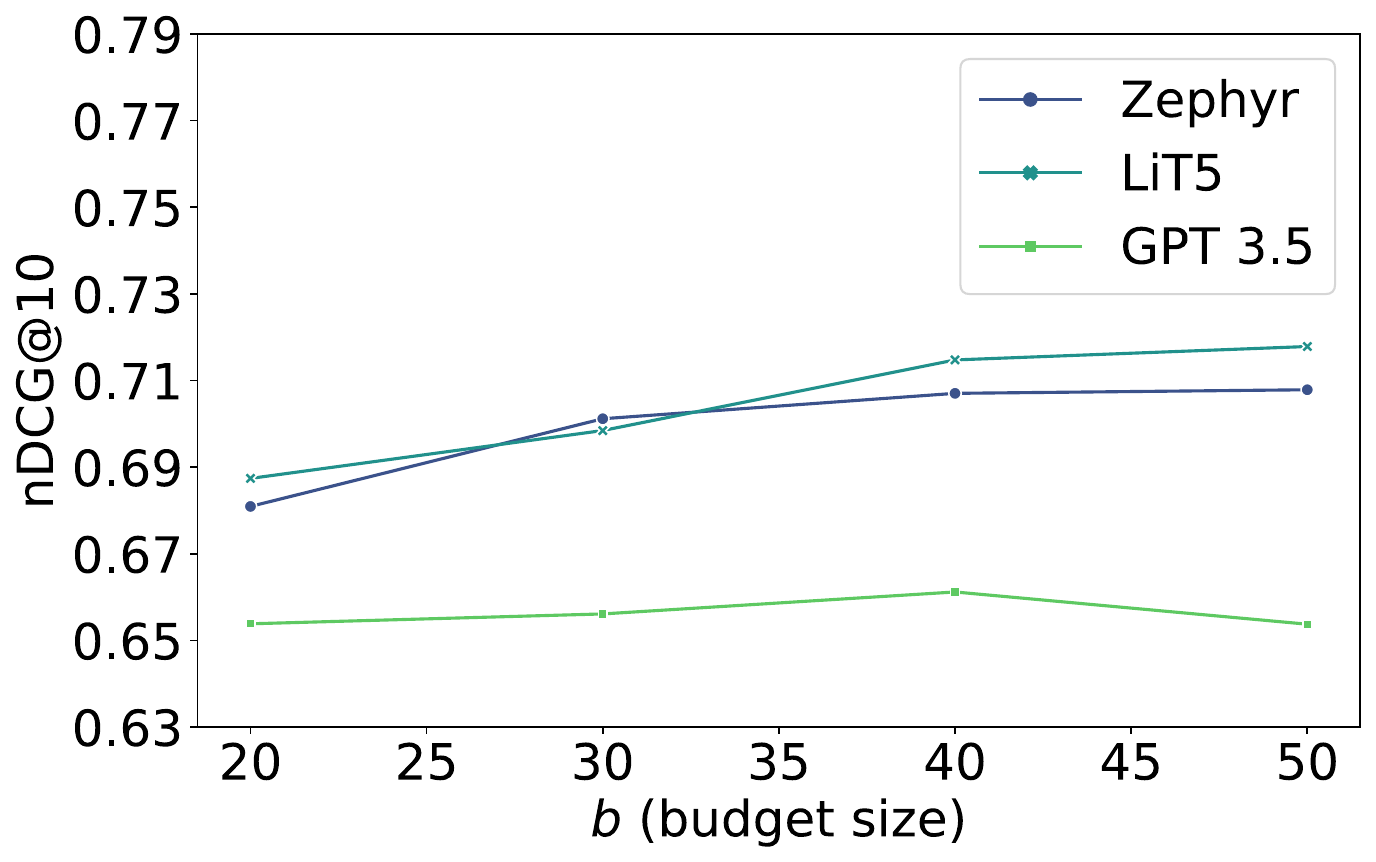}}
    \caption{Ablation of Budget across each first-stage ranker on the TREC Deep Learning 2019 test queries.}
    \label{fig:ablation}
\end{figure}

\para{Ablation of Budget.}
Addressing \textbf{RQ-4}, we conduct an ablation of our approach to assess how a larger budget and, therefore, the opportunity to re-rank the top-k multiple times can affect retrieval performance. We assess budgets in the range [20, 30, 40, 50]. In Figure \ref{fig:ablation}, observe that for precise first stages, small improvements in effectiveness occur when the budget is increased apart from RankGPT, which exhibits high variance. Additionally, we find that larger values of the budget parameter can exceed reported performance on Deep Learning queries from prior works; however, we do not tune this value on a validation set, and any improvement is insignificant over previous state-of-the-art methods. Akin to the multiple re-ranking stages proposed by ~\citet{rankzephyr} in which the previous ranking by the list-wise ranker is again re-ranked; however, we find a budget increase somewhat unnecessary given a first-stage dense or sparse neural retriever as performance gains are insufficient compared to the increased number of inferences required for a larger budget. However, in the case of BM25, we see that increasing the budget improves performance across all models, particularly in the case of LiT5, where nDCG@10 is improved from 0.687 to 0.715 when the budget is increased. Generally, we observe improvements of around 2 points of nDCG@10, allowing RankGPT to reach the performance of LiT5 when re-ranking SPLADE++ ED. We attribute this to the low precision of BM25, meaning that a greater budget is required to recover from a poor initial pivot value such that a progressive re-ranking is helpful. This explains small albeit insignificant degradation in nDCG@10 performance compared to a sliding window in Table \ref{tab:main}. Under our algorithm, progressive re-ranking does not require full re-ranking and is instead adaptive to the initial candidate set, improving efficiency.

\subsection{Out-of-Domain Effectiveness}

\begin{table}[tp]
\caption{Comparing `Single Window', `Sliding Window', and `TDPart' (ours) approaches on out-of-domain benchmarks. We report TOST (p<0.05), equivalence to our approach denoted with \underline{underline}. We report mean inferences with the mean number of inferences run in parallel in parentheses.
}

\begin{adjustbox}{width=\textwidth}
\begin{tabular}{@{}llrrrrrrrrr@{}}
\toprule
& & \multicolumn{4}{c}{TREC COVID} & \multicolumn{4}{c}{Touche} & \\
\cmidrule(r){3-6}
\cmidrule(r){7-10}
%\cmidrule{11-11}
Model & Mode & nDCG@1 & nDCG@5 & nDCG@10 & P@10 & nDCG@1 & nDCG@5 & nDCG@10 & P@10 & N. Inf. \\
\cmidrule(r){1-2}
\cmidrule(r){3-6}
\cmidrule(r){7-10}
\cmidrule{11-11}
Oracle & Single & 0.980 & 0.946 & 0.874 & 0.778 & 0.959 & 0.814 & 0.615 & 0.402 & 1.0 (1.0) \\
Oracle & Sliding & \underline{1.000} & \underline{0.989} & \underline{0.983} & \underline{0.974}  & \underline{1.000} & \underline{0.977} & \underline{0.877} & \underline{0.673} & 9.0 (1.0) \\
Oracle & TDPart & 1.000 & 0.989 & 0.975 & 0.954 & 1.000 & 0.977 & 0.876 & 0.671 & 6.5 (4.5) \\
\midrule
Zephyr & Single & 0.890 & 0.813 & 0.754 & 0.632 & 0.480 & 0.383 & 0.356 & 0.265 & 1.0 (1.0) \\
Zephyr & Sliding & 0.870 & 0.842 & \underline{0.805} & 0.702 & 0.490 & 0.342 & 0.318 & 0.222 & 9.0 (1.0) \\
Zephyr & TDPart & 0.910 & 0.854 & 0.804 & 0.696 & 0.480 & 0.369 & 0.340 & 0.241 & 5.8 (3.8) \\
\midrule
LiT5 & Single & 0.820 & 0.754 & \underline{0.715} & 0.608 & 0.429 & 0.366 & 0.340 & 0.255 & 1.0 (1.0) \\
LiT5 & Sliding & 0.850 & 0.763 & 0.744 & 0.644 & 0.439 & 0.330 & 0.323 & 0.239 & 9.0 (1.0) \\
LiT5 & TDPart & 0.820 & 0.769 & 0.725 & 0.618 & 0.418 & 0.337 & 0.331 & 0.251 & 5.2 (3.2) \\
\midrule
GPT 3.5 & Single & 0.880 & 0.779 & 0.718 & 0.598 & 0.459 & 0.374 & 0.325 & 0.237 & 1.0 (1.0) \\
GPT 3.5 & Sliding & 0.910 & 0.833 & 0.774 & 0.656 & 0.357 & 0.322 & 0.308 & 0.235 & 9.0 (1.0) \\
GPT 3.5 & TDPart & 0.900 & 0.849 & 0.784 & 0.682 & 0.276 & 0.308 & 0.272 & 0.198 & 6.5 (4.5) \\
\bottomrule
\end{tabular}
\end{adjustbox}
\label{tab:beir}
\end{table}

Table \ref{tab:beir} presents the out-of-domain performance of list-wise rankers using different partitioning algorithms. We observe again that our approach is statistically equivalent to a sliding window approach under an oracle setting but note that unlike MSMARCO, a single window is generally insufficient in terms of recall. Our approach improves over a single window on TREC COVID in 2 of 3 models and, on Touche, outperforms a sliding window approach in terms of nDCG@10 and P@10 using both RankZephyr and LiT5 whilst reducing inferences by over 33\%. We observe reduced performance assessed over TREC COVID using LiT5 compared to a sliding window approach, potentially showing that given the large training process of such rankers, without interventions (such as those applied in RankZephyr, e.g. order shuffling) to distinguish non-relevant documents from a pivot is more challenging. For Touche, both a sliding window and our approach generally degrade over a single window suggesting that these models cannot separate relevant and non-relevant documents when searching to a greater depth, this effect is most pronounced in RankGPT, which presents large degradation in terms of nDCG@10. Though our approach is a step towards more robust ranking using list-wise rankers, performance in contrast to a single window is still reduced, and such cases should be studied further to compensate for domain shift. The lack of statistical equivalence between approaches over list-wise rankers would suggest that greater work is required to improve their robustness and, therefore, allow greater recall. We note a clear trend that stronger models are more effective when applying a pivot-based approach and generally outperform the exhaustive sliding window while providing a sufficient trade-off between performance and efficiency.

\section{Conclusion}

In applying principles from dynamic pruning and selection algorithms, we have introduced a partitioning algorithm for list-wise ranking, which, while improving efficiency, matches or exceeds prior work in and out-of-domain. In investigating the precision of such rankers agnostic of a particular partitioning algorithm, we further motivate our approach using a pivot element chosen from highly ranked documents. Using a pivot as opposed to a strided window, we can rank documents to an arbitrary depth with a reduced number of inferences and in parallel. This finding is important in improving the efficiency of list-wise approaches as they find applications in both ranking and data annotation for more efficient rankers. Our proposed approach is statically equivalent to prior work in an oracle setting, suggesting that as underlying models improve, our algorithm can continue to improve efficiency with no performance degradation. Our approach is most effective with strong first-stage retrievers, where a single high-quality pivot element can be used for efficient candidate selection. In contrast, weak first stages might require a larger budget allowed by our algorithm for progressive re-ranking to overcome the initial imprecision. We identified limitations in out-of-domain retrieval tasks, where discrimination between relevant and non-relevant documents becomes more challenging at greater re-ranking depths. This highlights the need for further research on improving the robustness of list-wise rankers, particularly for generalizability across different domains. In summary, we show that current approaches are sub-optimal, and further consideration must be given to the efficiency of list-wise ranking at an algorithmic level.

\bibliography{sample-ceur}

\end{document}